# Investigations of key issues on the reproducibility of high-$T_c$ superconductivity emerging from compressed $La_3Ni_2O_7$

Yazhou Zhou[1*], Jing Guo[1]*, Shu Cai[2]*, Hualei Sun[3]*, Chengyu Li[2], Jinyu Zhao[2], Pengyu Wang[1,4], Jinyu Han[1,4], Xintian Chen[1,4], Yongjin Chen[2], Qi Wu[1], Yang Ding[2], Tao Xiang[1,4,5], Ho-kwang Mao[2]† and Liling Sun[1,2,4]†

[1]*Institute of Physics, Chinese Academy of Sciences, Beijing 100190, China*

[2] *Center for High Pressure Science & Technology Advanced Research, Beijing 100094, China*

[3]*School of Science, Sun Yat-Sen University, Shenzhen, Guangdong 518107, China*

[4]*University of Chinese Academy of Sciences, Beijing 100190, China*

[5]*Beijing Academy of Quantum Information Sciences, Beijing 100193, China*

Recently, the signatures of superconductivity near 80 K have been discovered in the single crystal of $La_3Ni_2O_7$ under pressure, which makes it a new candidate of the high-temperature superconductors dominated by 3*d* transition elements after the cuprate and iron-pnictide superconductors. However, there are several critical questions that have been perplexing the scientific community. These questions include (1) what factors contribute to the inconsistent reproducibility of the experimental results? (2) what is the fundamental nature of pressure-induced superconductivity: bulk or non-bulk (filamentary-like)? (3) where does the superconducting phase locate within the sample if it is filamentary-like? (4) is the oxygen content important for developing and stabilizing its superconductivity? In this study, we employ comprehensive high-pressure techniques to address these crucial issues. Through our modulated *ac* susceptibility measurements, we are the first to find that the superconductivity in this nickelate is filamentary-like. Our scanning transmission electron microscopy (STEM) investigations suggest that the filamentary-like superconductivity most likely emerges at the interface between the $La_3Ni_2O_7$ and $La_4Ni_3O_{10}$ phases. By tuning the oxygen content on the polycrystalline $La_3Ni_2O_7$, we also find that the oxygen content plays vital role for developing and stabilizing its superconductivity. The upper and lower bounds of the oxygen content are 7.35 and 6.89, respectively. Our results provide not only new insights into understanding the puzzling issues in this material, but also significant information for achieving a better understanding on the superconductivity of this material.

The quest for novel high-temperature superconductors is motivated by the expectation that these materials could have practical applications or lead to a better understanding of the fundamental mechanisms underlying high-temperature superconductivity. A possible fruitful area predicted to investigate is the nickelates[1]. Several candidates for superconductivity have been found in the thin-film nickelates[2–6]. Recently, the signatures of superconductivity with a transition temperature ($T_c$) near 80 K have been discovered in the single crystal $La_3Ni_2O_7$ under high pressure[7]. This high-$T_c$ compound has garnered considerable interest in the past months, and a great deal of advancements have been achieved primarily [8–50]. The results from the high-resolution angle-resolved photoemission spectroscopy[21,51,52] suggest that the ambient-pressure $La_3Ni_2O_7$ exhibits a strong correlation characeristic[13] associated with the covalent hybridization between transition metal 3*d* and O 2*p* orbitals, which is similar to the formation of the Zhang-Rice singlet in cuprates[53]. In particular, the coexistence of spin-density wave and charge-density wave phases has been revealed at ambient pressure [20, 23]. These observations suggest that $La_3Ni_2O_7$ holds the potential to exhibit intriguing phenomena under specific non-thermal control parameters, such as pressure.

On the experimental side, although zero resistance has been observed in some compressed nickelate samples[8,9], most of them cut even from the same batch show no zero resistance superconductivity. The anomalously irreproducible superconductivity suggests the possibility of an inhomogeneous distribution of a small amount of the superconducting phase within the crystal. In addition, the signature of the high-$T_c$ superconductivity was also observed in the $La_3Ni_2O_7$ polycrystalline samples, but no Meissner effect was reported in these papers[9,10,11]. Recent experimental results regarding the characterization of the $La_3Ni_2O_7$ single crystals have revealed the existence of minor intergrowth phases within the main phase characterized by bilayer blocks with alternating $NiO_6$ octahedra (referred to as 2222 or 327 phase). These intergrowth phases include the 214 phase characterized by single layer blocks, the 4310 phase characterized by trilayer blocks, and a polymorphic phase featuring both single and trilayer blocks (referred to as 1313 phase) [12,14,16, 22,26]. These results suggest that this material may exhibit a more complex physics associated with high-$T_c$ superconductivity compared to the pure 327 phase, which needs a comprehensive investigation.

In this study, we conducted the first high-pressure measurements on the superconducting volume fraction of the compressed $La_3Ni_2O_7$ using a modulated alternating-current (*ac*) susceptibility system. The employed system possesses the ability to detect the superconducting volume fraction as low as 0.5% (the details can be found in the Supplementary Information - SI). Moreover, we performed *in-situ* resistance, Hall coefficient and scanning transmission electron microscopy (STEM) investigations, with the attempt to elucidate the underlying issues of the anomalous superconductivity observed in the compressed $La_3Ni_2O_7$.

Before conducting the high-pressure experiments, we performed structural characterization and resistance measurements on the single crystal $La_3Ni_2O_7$ sample at ambient pressure (see SI). The single crystal X-ray diffraction reveals that the structure of our sample crystalizes in an orthorhombic unit cell in the *Amam* space, in good agreement with the results reported in Ref. 7. Careful examination of our XRD data reveals the presence of the 1313 phase in one of the two single crystal samples (see SI), which aligns with the findings reported by Chen *et al*[12]. The temperature dependence of resistance for the ambient-pressure sample shows a metallic behavior upon cooling, consistent with the reported results[7]. In the high-pressure modulated *ac* susceptibility measurements, we loaded a piece of elemental vanadium, with the identical shape but nearly half volume of the $La_3Ni_2O_7$ sample, in the same high-pressure chamber as a reference to estimate the relative superconducting volume fraction of the $La_3Ni_2O_7$ sample. By comparing their diamagnetic signal magnitude, we provide a rough estimation for the superconducting volume fraction of the $La_3Ni_2O_7$ sample. The pressure applied on the sample was determined through the pressure-dependent $T_c$ of the vanadium[54]. Figure 1 shows the temperature dependence of the modulated *ac* susceptibility ($\Delta\chi'$) obtained at different pressures. Below 20.7 GPa, no superconducting transition is observed, and only the background signal is detected (Fig. 1a-1c). While, in this pressure range, the superconducting transitions of the elemental vanadium can be clearly identified by a sudden increase in the signal above the background (the arrow indicates the onset temperature of the superconducting transition, see the insets of Fig. 1a-1c). Upon increasing the pressure to 22 GPa, a 'broad-peak' emerges at 64.3 K (Fig. 1d). According to the theory of the modulated magnetic susceptibility[55–57], this 'peak' results from the superconducting transition of the sample, instead of a magnetic transition (see SI). To enhance the visibility, we mark the broad peak in light green with the background represented by a dashed line. This kind of broad peak can be observed up to 28.2 GPa, the highest pressure investigated in this run (Fig.

1d-1f).

Here, we employed the ratio of the superconducting diamagnetic signal per volume of the sample ($\Delta\chi_{s}'/V_s$) to that of the vanadium ($\Delta\chi_{v}'/V_v$) to estimate the relative superconducting volume fraction in $La_3Ni_2O_7$, yielding a value of 0.68% (~1%) $[(\Delta\chi_{s}'/V_s)/(\Delta\chi_{v}'/V_v)]=(0.011/16.6926\times10^4(\mu m)^3)/(0.524/5.4\times10^4(\mu m)^3)]$, where $V_s=93\times93\times19.3= 16.6926\times10^4(\mu m^3)$ and $V_v=75\times75\times9.6= 5.4\times10^4(\mu m^3)$. This finding suggests that the superconductivity observed in $La_3Ni_2O_7$ is non-bulk nature but a filamentary-like one.

Furthermore, we found that the superconducting transition temperature slightly decreases with increasing pressure, as shown in Fig. 1d-1f, consistent with the resistance measurements reported in the previous studies[7-9]. Then, we released the pressure starting from 28.2 GPa and found a slight increase in $T_c$ (Fig.1g-1h), in which the maximum $T_c$ reached 65.2 K at 21.2 GPa. As the pressure was further released to 17.2 GPa, the superconducting diamagnetic signal became invisible, consistent with the behavior where the pressure was increased to the same level (Fig.1b and 1c). This indicates that the pressure-induced transition is reversible. However, unexpectedly, in our five independent measurements on the samples cut from the same batch, we observed the superconducting diamagnetic transition only once (see SI). In addition, we also did the *ac* susceptibility measurement on the $La_3Ni_2O_7$ single crystal sample using an unmodulated *ac* susceptibility system (the sample signal is detected by a single-phase-locked amplifier), did not observe any evidence of superconducting diamagnetic transition (see SI).

To further investigate the superconducting behavior on the compressed $La_3Ni_2O_7$ single crystal, we conducted *dc* susceptibility measurements on this material and superconducting vanadium (Fig.2). No evidence of superconducting transition was found in $La_3Ni_2O_7$, although the volume of the $La_3Ni_2O_7$ single crystal sample is nearly two times larger than that of the vanadium. These results suggest that the $La_3Ni_2O_7$ single crystal doesn't exhibit bulk superconductivity.

Next, we performed high-pressure resistance measurements on the $La_3Ni_2O_7$ single crystal samples. We have measured seven samples but observed the presence of superconductivity with zero resistance only in one of them. Figure 3 shows the results from two of the samples (A and B). Sample A displays a semiconducting-like behavior at a pressure below 7.9 GPa. Its resistance starts to drop at the low temperature at 11.1 GPa (Fig. 3a). This drop becomes more pronounced with increasing pressure. It approaches zero above 21.5 GPa (Fig. 3b). For sample B, zero resistance is observed when the pressure is applied between 17.8 GPa and 31.5 GPa (see the inset of Fig. 3c), which confirms the occurrence of superconducting transition in $La_3Ni_2O_7$. Subsequently, we released the pressure from 31.5 GPa down to 5.5 GPa and observed a gradual loss of superconductivity (Fig. 3d), consistent with our findings from the modulated *ac* susceptibility measurements (Fig. 1). At 5.5 GPa, the sample returns to its semiconducting phase (Fig. 3d), indicating that the pressure-induced transition from the semiconducting state to the superconducting state is reversible.

Figure 4 shows the phase diagram obtained based on our *ac* susceptibility and transport measurements on $La_3Ni_2O_7$. In this diagram, we have also included the density-wave-like phase transition temperatures ($T_{SDW}$ and $T_D$) reported in Ref. 13,23 and the $T_c$s reported by other groups[7-9]. It is shown that the application of pressure suppresses the formation temperature of DW-like phase but enhances the formation temperature of the SDW phase [13,23], and then induces a superconducting (SC) transition at a critical pressure ($P_c$) above 10 GPa (Fig.4a).

In addition, we conducted Hall measurements on the sample (see SI) and found that the Hall coefficient $R_H$ remains positive across the investigated pressure range with a notable drop at around $P_c$ (Fig.4b). This positive Hall coefficient and the sudden change align with the trends observed in the studies on the Sr-doped $LaNiO_2$[3]. Since the sample undergoes a structure phase transition from an ambient-pressure orthorhombic *Amam* phase to a high-pressure orthorhombic *Fmmm* phase at the temperature of our Hall measurements[7,19], the drop in $R_H$ at the boundary between the DW-like and the normal state of the superconducting (SC) phases seems to be related to the pressure-induced structure phase transition. On the other hand, this change in charge carriers at the pressure around the emergence of superconductivity suggests that the pressure-induced change in oxygen vacancies or Ni valence states may also play a role in developing superconductivity in this material, which deserves further investigation.

The observation of small superconducting volume fraction and the fact that it is difficult to obtain a zero-resistance state suggest that the superconductivity in $La_3Ni_2O_7$ is likely filamentary in nature[17,58–61]. This speculation accounts for several enigmas regarding the pressure-induced superconductivity in the $La_3Ni_2O_7$ single crystal sample.

As an example, when conducting resistance measurements along different directions of the sample, we found that zero resistance was observed in only one direction while not in the other, despite having the same onset $T_c$ values in both directions.

In order to provide more evidence for the proposed filamentary-like superconductivity in compressed $La_3Ni_2O_7$, we performed high-pressure resistance measurements on the sample using various current values (ranging from 0.01 mA to 1 mA). We found a non-Ohmic behavior below $T_c$, which is consistent with the results reported recently[22], and non-linear *V-I* characteristic (see SI), supporting that the superconductivity in $La_3Ni_2O_7$ is non-bulk in nature, instead, a filamentary-like.

To further understand the perplexing superconducting behavior of $La_3Ni_2O_7$, we performed STEM investigations on both single crystal and polycrystalline samples (Fig. 5). It is found that their predominant phase is the $La_3Ni_2O_7$ (327) phase with the 2222-stacking sequence, while the minor phases (< 5%) consist of the $La_4Ni_3O_{10}$ (4310) and $La_2Ni_1O_4$ (214) phases in the polycrystalline sample (Fig.5a and 5b) and the 4310 phase in the single crystal samples (Fig.5c-5t). Furthermore, interfaces were observed within both the superconducting polycrystalline sample (Fig.5a) and the single crystal sample (Fig.5c-5n). Careful inspiration on the superconducting single crystal sample reveals the existence of the 327, 4310 and their mixed phases at the interfaces (Fig.5o-5q). By performing fast Fourier transforms (FFT) based on the corresponding electron diffraction results (Fig. 5r-5t), we have obtained the averaged *d*-spacing value profiles of the 327, 4310 and their mixed phases at the interface (Fig.5u). These results indicate that $La_3Ni_2O_7$ exhibits significant inhomogeneity.

Similar to what has been observed in cuprate superconductors[62–65], the deviation from stoichiometric oxygen content in $La_3Ni_2O_7$ appears to have a significant impact on the presence of superconductivity[17]. To establish the connection of the oxygen content with the presence of superconductivity in $La_3Ni_2O_7$, we performed comprehensive investigations starting with the synthesis of polycrystalline samples by using sol-gel method (see SI). The characterization on the as-grown samples found that they can be well indexed by the orthorhombic structure in the *Amam* space group (see SI), which is consistent with the structure of the single crystal sample reported in Ref. 7. The oxygen content of the as-grown sample was estimated to be about 6.88 determined by thermalgravimetric analysis (see SI). Subsequently, we adjusted the oxygen content over a wide range using the electrochemical method[66–67] (see SI), and conducted the measurements of resistance and modulated *ac* susceptibility on the same sample in a diamond anvil cell. As shown in Fig. 6a, the superconductivity was observed only within a specific range of the oxygen content, ranging from 6.88 ($\delta$ = -0.12) to 7.42 ($\delta$ = 0.42). Intriguingly, within this range, the maximum $T_c$ value remains nearly unchanged, which suggests that, when the oxygen content falls in an appropriate range, it no longer significantly affects the maximum $T_c$ value. Based on these results, we determined the upper and lower bounds of the oxygen content required for the development of superconductivity to be about 7.35 and 6.89, respectively. Outside this range, superconductivity was no longer observed.

Remarkably, we noticed that there exists a distinct narrow range of oxygen content inside the superconducting regime (see the shaded area in Fig.6). Within this regime, the oxygen content in $La_3Ni_2O_7$ closely approaches 7, and the corresponding normal-state resistance of the sample exhibits a metallic behavior. Differently, outside this regime the normal-state resistance demonstrates a semiconducting-like behavior, and these is a prominent upturn at low temperature. These observations indicate the crucial role of the stoichiometric oxygen content on the presence of metallic normal state and the superconducting state at low temperature in this material.

The modulated *ac* susceptibility measurements were also conducted on the same polycrystalline samples that we investigated for the resistance measurements. Not surprisingly, we did not observe any superconducting diamagnetic signal in the samples with oxygen contents ranging from 6.88 to 7.42 (see Fig. 6b to Fig. 6e). As a result, we propose that the superconducting volume fraction in these polycrystalline samples should be less than 0.5%, which is the detectable limit of our measuring system (see SI).

Based on these observations, we have the following discussions to address various perplexing issues:

(1) Despite the dominating phase of the single crystal sample is the 327 phase, the 327 phase itself should not be responsible for the superconductivity because only ~1% superconducting volume fraction has been observed in the material.

(2) Due to the lower $T_c$ value (~20 K) of the 4310-sample compared to that of the 327-sample, it is implausible that the observed superconducting transition near 80 K observed in $La_3Ni_2O_7$ arises from the pure 4310 phase.

(3) The commonness of the coexistence of the 327 and 4310 phases in both single

crystal and polycrystalline samples leads us to propose that the filamentary-like superconductivity is most likely located at the interface between the 327 and 4310 phases.

(4) Although we detected a polymorph with a 1313 stacking sequence in one of the single crystal samples using single crystal X-ray diffraction (see SI), our STEM investigations of both single crystal and polycrystalline samples did not find the presence of this polyform. The absence of the 1313 phase in the samples of our STEM investigations suggests that the $La_3Ni_2O_7$ samples exhibit significant inhomogeneity. Further investigations are imperative to explore whether filamentary superconductivity is linked to the 1313 phase. It is important to highlight that the 4310 phase from the outer layer of the 1313 phases can also create an interface between the 327 and 4310 phases, which align with our proposed possibility regarding the potential location of filamentary superconductivity, specifically at the interface between the 327 and 4310 phases.

In conclusion, we are the first to confirm the existence of the high-temperature superconductivity in $La_3Ni_2O_7$ through demonstrating the zero resistance and superconducting diamagnetism, and subsequently propose that the nature of the observed superconductivity is filamentary-like. These findings provide vital insights in understanding the anomalous superconducting behavior of the compressed $La_3Ni_2O_7$. Furthermore, our measurements of resistance and *ac* susceptibility consistently demonstrate the reversibility of pressure-induced superconductivity, indicating that the superconducting phase exists in a metastable state. Our STEM investigation on both of the $La_3Ni_2O_7$ single crystal and polycrystalline samples reveal that the observed filamentary-like superconductivity in this material most likely emerges at the interface between 327 and 4310 phases. If this is case, it is reasonable to anticipate that enhancing the interface between the 327 and 4310 phases is a potential route for increasing the superconducting volume fraction[68]. Further, we have determined the upper and lower bounds of the oxygen content for the presence of superconductivity in the compressed $La_3Ni_2O_7$, approximately 7.35 and 6.89, respectively.

Our findings then raise significant questions that warrant further investigation, which include (1) Why does the coexistence between the 327 and 4310 phases possess the ability to develop the high-$T_c$ superconductivity? (2) Is the valence change of Ni ions in the compressed $La_3Ni_2O_7$ related to its superconductivity? (3) Why the stoichiometric oxygen content is favorable for the presence of superconductivity? Addressing these questions may not only provide deeper insights into understanding the underlying physics for the superconductivity in this nickelates, but also give an effective guidance for exploring new high-$T_c$ superconductors in *3d* transition metal compounds.

## Acknowledgements

The work was supported by the National Key Research and Development Program of China (Grant No. 2022YFA1403900 and 2021YFA1401800), the NSF of China (Grant Numbers Grants No. U2032214, 12122414, 12104487 and 12004419) and the Strategic Priority Research Program (B) of the Chinese Academy of Sciences (Grant No. XDB25000000). J. G. is grateful for supports from the Youth Innovation Promotion Association of the CAS (2019008). This work was supported by the Synergetic Extreme Condition User Facility (SECUF).

## Author contributions

L.S., H.K.M., T.X. and Q.W. designed the study and supervised the project. H.L.S. provides the $La_3Ni_2O_7$ single crystals. Z.Y.Z. synthesized the $La_3Ni_2O_{7-\delta}$ and $La_3Ni_2O_{7+\delta}$ polycrystalline samples. Z.Y.Z. and S.C. performed high-pressure modulated *ac* susceptibility measurements, J.G., S.C., P.Y.W., J.Y.Z. and J.Y.H. performed the high-pressure resistance and Hall coefficient measurements. Z.Y.Z. performed experiments on electrochemical reaction. S.C. conducted single crystal x-ray diffraction measurements for the ambient-pressure sample. C.Y.L. and Y.J.C. performed STEM investigations. L.S., T.X., Q.W., Y.D., H.K.M., Y.Z. Z., J. G. and S.C. analyzed data. L.S., H.K.M., T.X., Q.W., Y.Z.Y. and J.G. wrote the manuscript with the efforts of all authors.

## Additional information

These authors with star (*) contributed equally to this work.

Correspondence and requests for materials should be addressed to Ho-kwang Mao (maohk@hpstar.ac.cn) and Liling Sun (llsun@iphy.ac.cn or liling.sun@hpstar.ac.cn)

## Competing interests

The authors declare no competing interests.

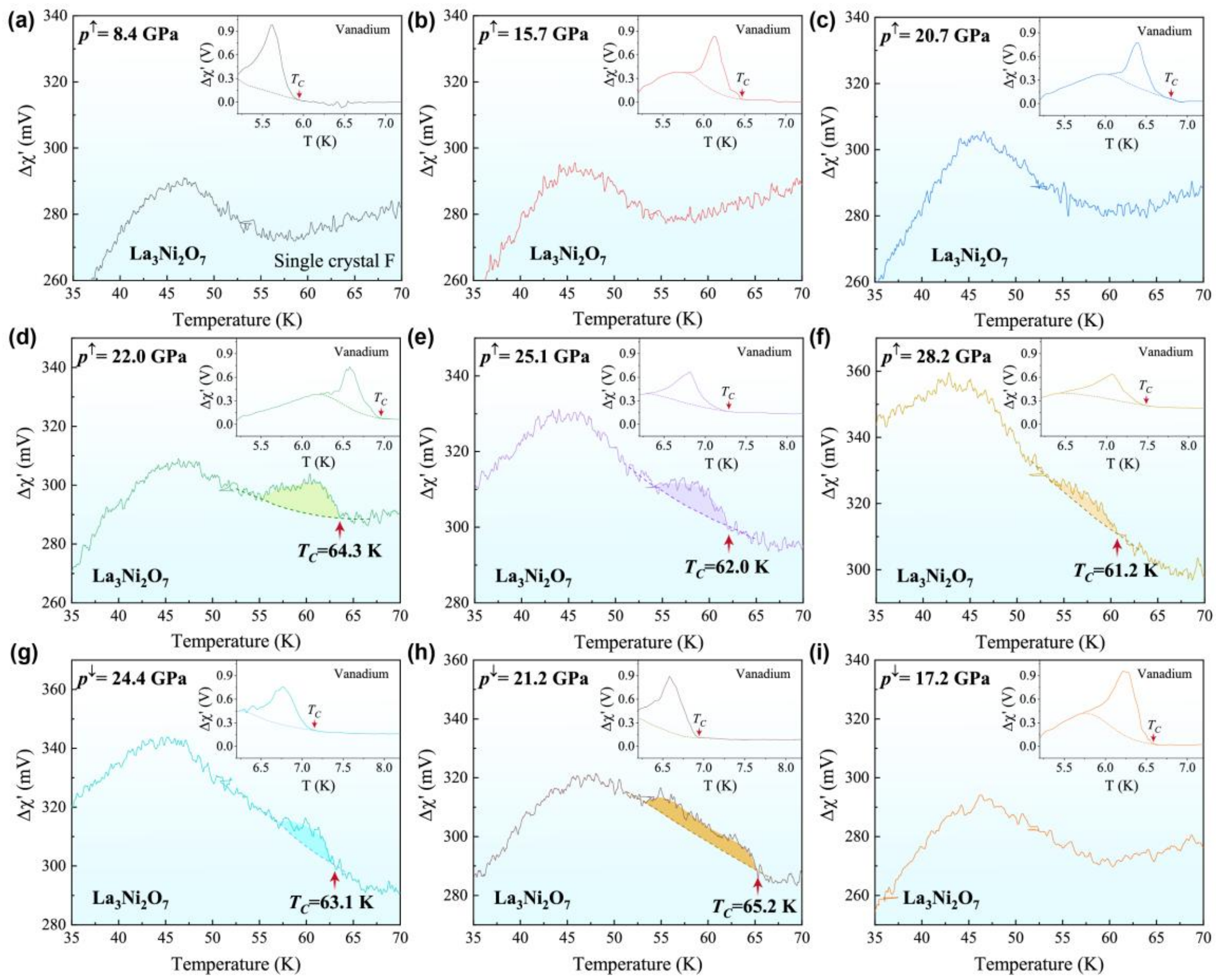

**Figure 1 The raw data of the modulated *ac* susceptibility (Δχ′) versus temperature**

**($T$) measured at different pressures for the $La_3Ni_2O_7$ single crystal.** (a-c) The results obtained at 8.4, 15.7 and 20.7 GPa, respectively. Upon compression, no superconducting diamagnetic signal is observed. (d-f) The plots of $\Delta\chi'$ as a function of $T$ measured at 22.0, 25.1 and 28.2 GPa, revealing a superconducting diamagnetic transition occurring at temperatures of 63.4, 62.0 and 61.2 K. (g-i) The data measured during the pressure releasing process, showing the loss of superconductivity at 17.2 GPa. The insets depict the corresponding superconducting transition of the elemental vanadium, captured through synchronous measurements with the $La_3Ni_2O_7$ single crystal in the same high-pressure chamber. The onset $T_c$ is indicated by the red arrow. By comparing the jump height of the sample and the vanadium, the relative superconducting volume fraction of $La_3Ni_2O_7$ is estimated to be 0.68% (approximately 1%) at 22.0 GPa, suggesting that the superconductivity of the compressed $La_3Ni_2O_7$ is filamentary-like.

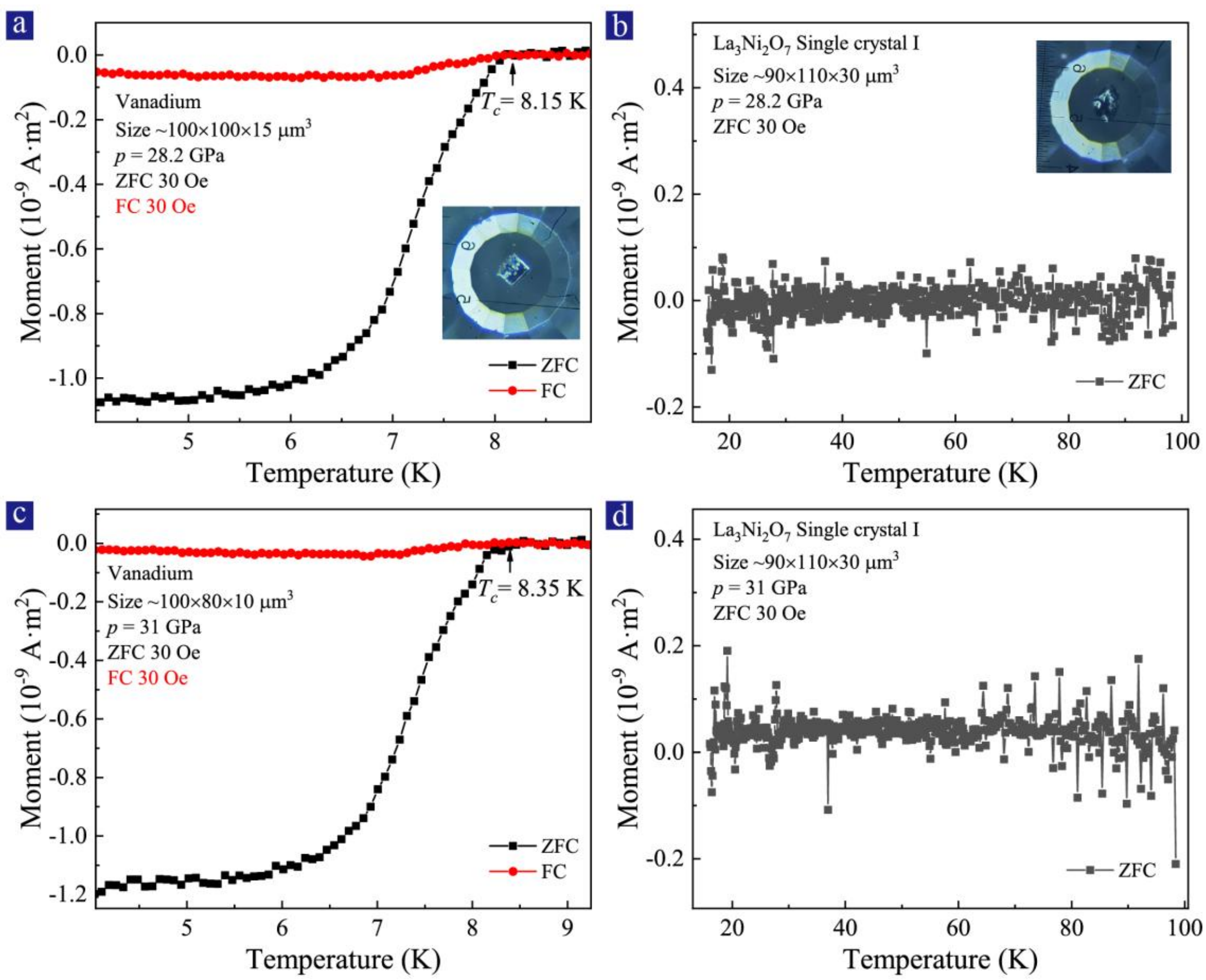

**Figure 2 The temperature dependence of *dc* susceptibilities for the superconducting vanadium and the $La_3Ni_2O_7$ single crystal.** No evidence of superconducting transition is observed in the $La_3Ni_2O_7$ single crystal at 28.2 GPa and 31 GPa, in the pressure range of which the superconductivity of $La_3Ni_2O_7$ should appear.

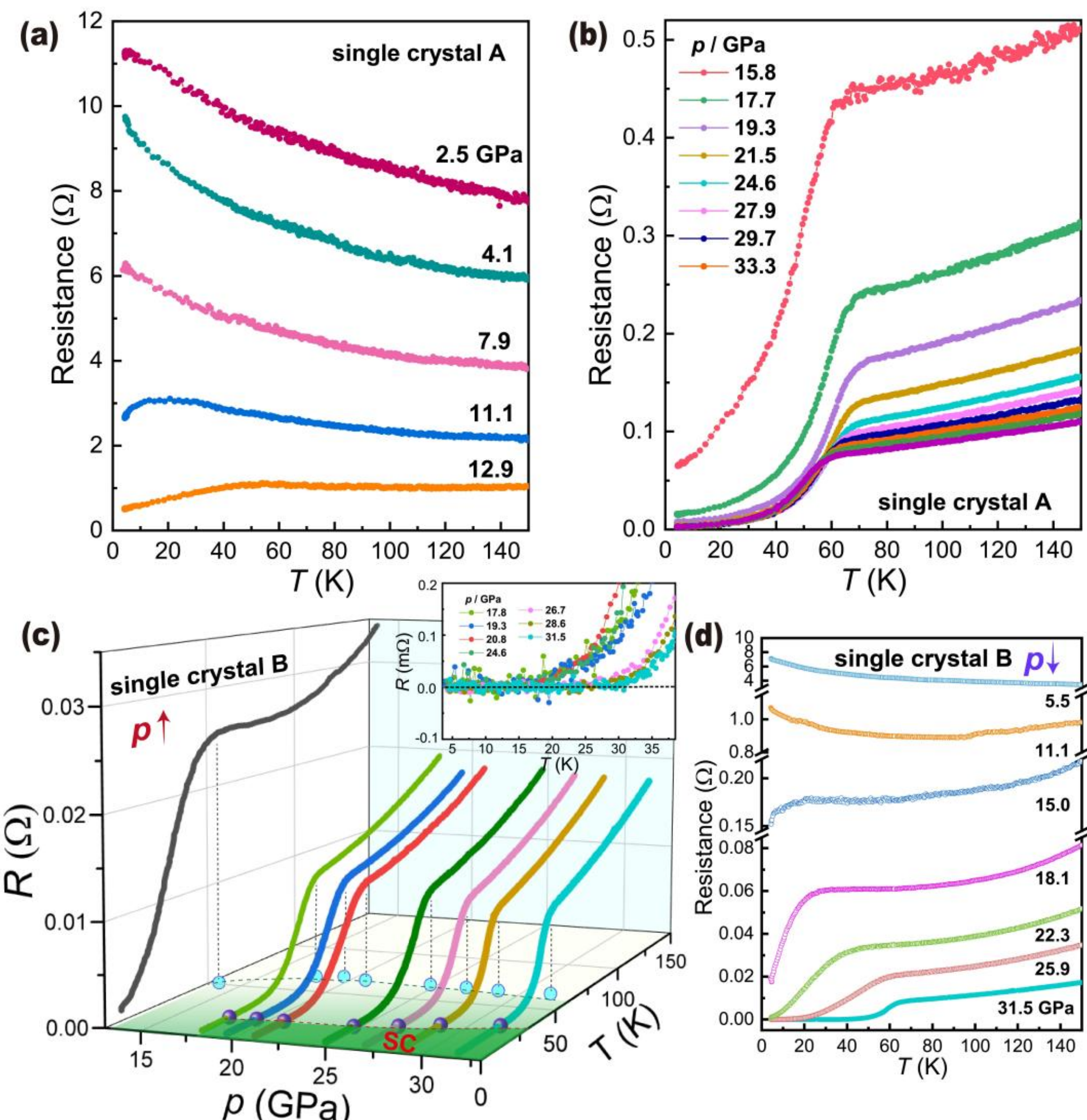

**Figure 3 The temperature dependence of resistance measured at different pressures for the $La_2Ni_3O_7$ single crystals.** (a-b) The results of resistance measurements on sample A within the pressure range of 2.5-33.3 GPa, illustrating the evolution from a semiconducting-like state to a superconducting-like state. (c) Resistance-temperature curves at different pressures for sample B, showing a superconducting transition with zero resistance in the pressure range of 17.8-31.5 GPa. The upper right panel displays an enlarged view of the low-temperature resistance. (d) The results obtained from the pressure release measurements, demonstrating a gradual disruption of the superconducting state.

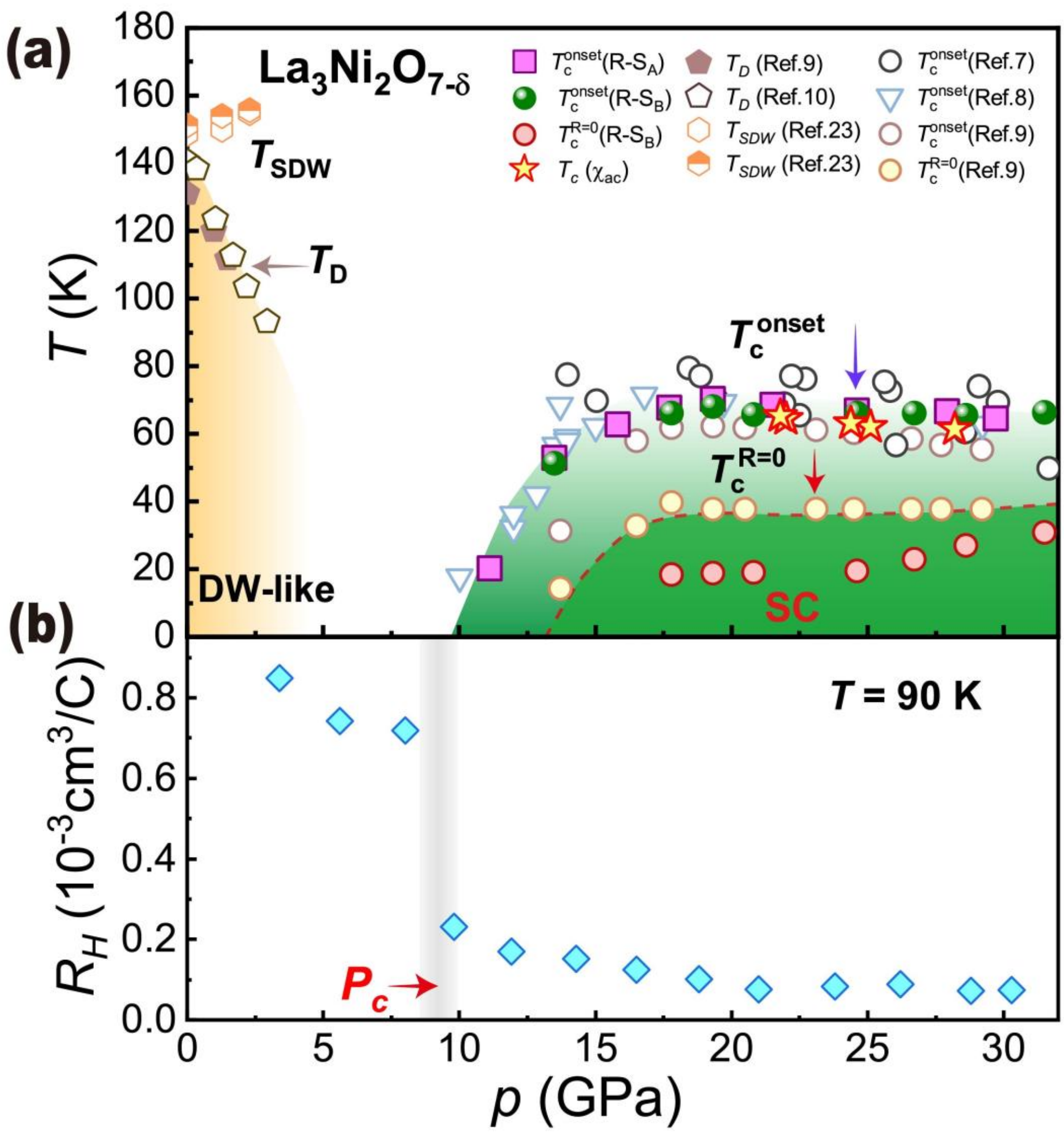

**Figure 4 The pressure-temperature phase diagram and Hall coefficient ($R_H$) as a function of pressure for the $La_3Ni_2O_7$ single crystal.** (a) A summary of our results and the reported results obtained from high-pressure modulated *ac* susceptibility and resistance measurements. The filled stars represent the data from our susceptibility measurements. The green balls, squares and circles filled with pink are the data from our resistance measurements. DW and SC stand for density-wave and superconducting phases, respectively. $T_D$ denotes the onset temperature of the DW-like phase transition, while $T_c^{onset}$ *and* $T_c^{R=0}$ represent the onset and zero resistance temperature of the superconducting transition, respectively. (b) The plot of pressure versus Hall coefficient ($R_H$) measured at 90 K, demonstrating a significant drop in $R_H$ around the boundary between the DW-like phase and the SC phase.

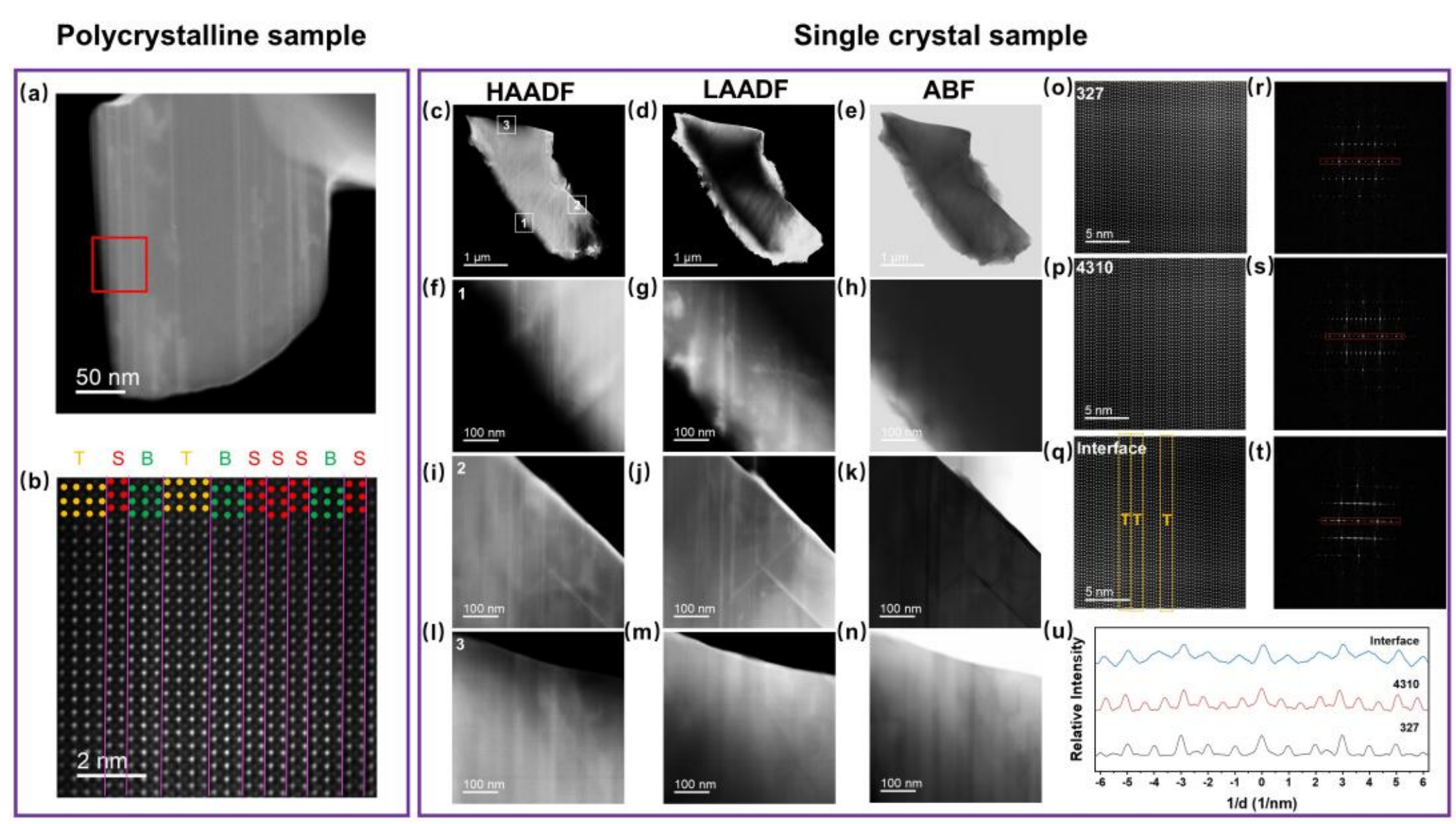

**Figure 5 The scanning transmission electron microscopy (STEM) images of the superconducting samples.** (a) Low magnification low-angle annular dark-field (LAADF) image along the [110] direction of the $La_3Ni_2O_{7.07}$ polycrystalline sample. The extensive gray regions correspond to the pure 327 phase with the 2222-stacking sequence, while the white line regions indicate the interface structure. (b) An atomic-scale high-angle annular dark-field (HAADF) image along the [110] direction obtained from the area within the red box in figure (a). This image reveals the layered stacking interface structure. The atoms of La in the 327, 4310, and 124 phases are represented by green, orange, and red dots, respectively. T, B, and S denote the trilayer, bilayer, and single-layer arrangements of Ni-O planes. (c), (d) and (e) Low magnification HAADF

image, LAADF image and annular bright field (ABF) image, along the [110] direction of the $La_3Ni_2O_7$ single crystal sample, which show many white lines (interface structures) existed in the single crystal (the sample size is about 3 μm). (f)-(n) The STEM images obtained from the area within the boxes labeled by 1, 2 and 3 in figure (c), (d) and (e), displaying many interfaces within the single crystal. (o-q) The atomic-scale images of the 327, 4310 phases and their mixed phase at the interfaces in the $La_3Ni_2O_7$ single crystal sample. (r-t) The corresponding fast Fourier transforms (FFT) diffraction patterns of the 327, 4310 phases and their mixed phase at the interfaces, and the diffuse streaks shown in (t). (u) The corresponding intensity profile of FFT diffraction patterns.

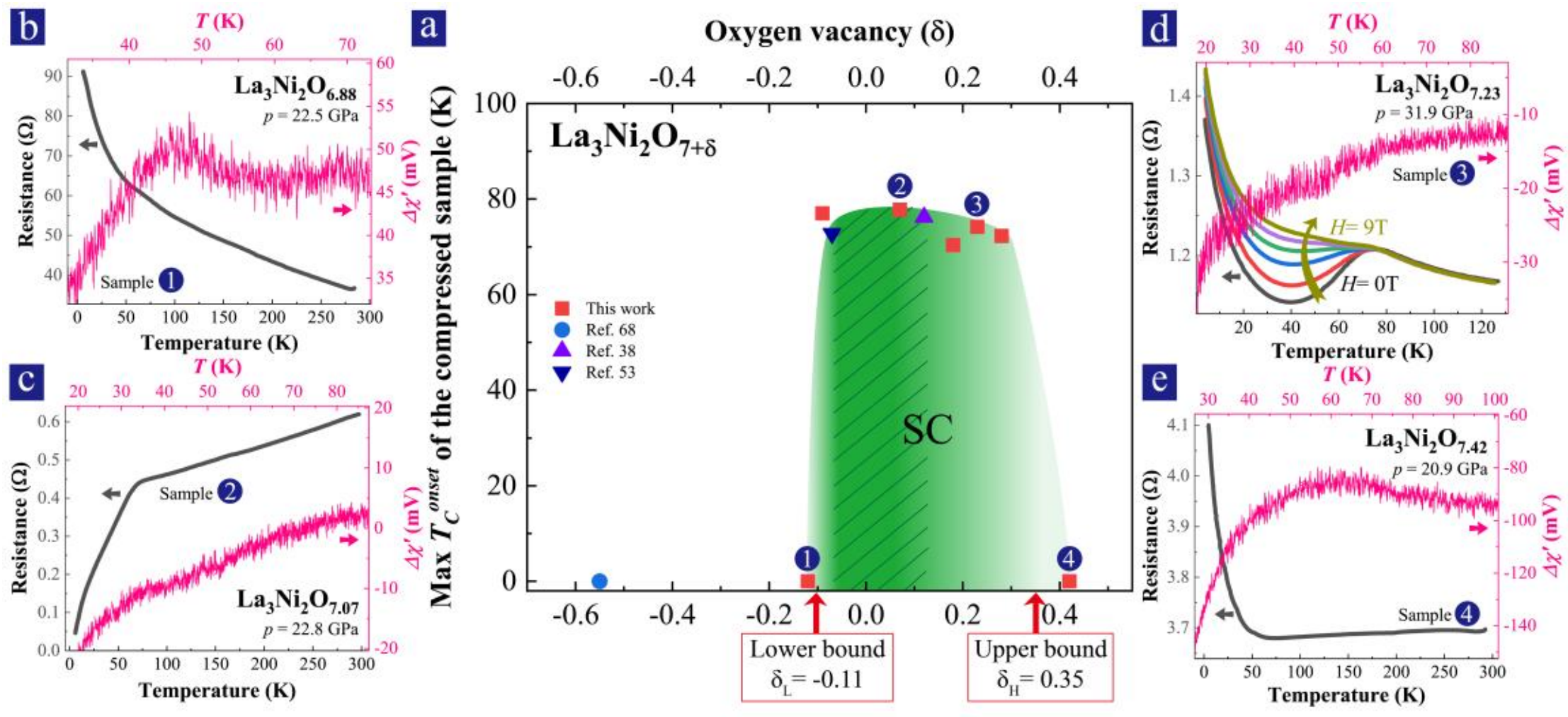

**Figure 6 The lower and upper bounds of the oxygen vacancy for the presence of superconductivity.** (a) The connection of the oxygen vacancy in $La_3Ni_2O_{7+\delta}$ with the superconductivity. The lower and upper bounds of the oxygen vacancy for the presence of superconductivity are estimated to be about -0.11 and 0.35 respectively, the value of which is taken by an average of oxygen content for non-superconducting and superconducting samples ($\delta_L$=[(7-6.88)+(7-6.91)]/2=0.11, $\delta_H$=[(7.42-7)+(7.28-7)]/2=0.35). The shaded region represents the range where the low temperature resistance of the sample exhibits metallic behavior, while outside of this region, the low temperature resistance of the sample displays a noticeable upturn. (b)-(e) The results of temperature-resistance and temperature-susceptibility for the samples with different oxygen contents. The combined *ac* susceptibility and resistance measurements were conducted in the two different diamond anvil cells (DAC #1 and DAC#2). We identified that the background of these two DACs is not the same. The modulated *ac* susceptibility measurements for sample #1 and sample #4 were conducted in DAC #1, resulting in both samples exhibiting the same background signal (Fig.6b and 6e). Whereas the susceptibility measurements for sample #2 and sample #3 were performed in DAC#2, leading to both samples displaying the same background signal (Fig.6c and 6d).